# Non-equilibrium THz-phonon spin coupling in CrI$_3$

V Shokeen[1], M Pavelka[1], R Chulkov[1], A Yaroslavtsev[1,2], J Rogvall[1], D Muradas Belinchón[1], U Noumbe[1], M Abdel-Hafiez[3], M. Venkata Kamalakar[1], O Grånäs[1] and H A Dürr[1]

[1]Department of Physics and Astronomy, Uppsala University, Uppsala, Sweden

[2]MAX IV Laboratory, Lund University, Lund, Sweden

[3]Department of Applied Physics and Astronomy, University of Sharjah, Sharjah, UAE

E-mail: vishal.shokeen@physics.uu.se

**Abstract.** Manipulating magnetism at the THz timescale in atomically thin ferromagnets by exploiting the interactions of spins with optical phonon modes presents an innovative idea for THz spintronics and magnonics. Utilizing the coupling of phonon modes to the magnetization could lead to new ways of generating and controlling spin wave excitations in future applications. We use femtosecond optical laser pulses to generate excitons, bound electron-hole pairs, in bulk-like ferromagnet CrI$_3$ flakes and probe the subsequent charge and spin dynamics with optical pump-probe spectroscopy. In CrI$_3$, exciton formation is known to drive coherent optical phonon modes with 2.4 and 3.9 THz frequencies corresponding to the bending and stretching of the Cr-I bonds. We show that both phonon modes also lead to magnetization oscillations. This establishes spin-phonon coupling for both modes, contrary to previous observations that only report the 3.9 THz phonon mode can influence the CrI$_3$ magnetization.

## 1. Introduction

The fundamental interaction between atomic structure and magnetic ordering provides a versatile tool to control the properties of materials. Manipulating electron spins via phonons or vice-versa opens the path for controlling magnetic or structural phase transitions. An important and pertinent question is the mechanism for the conservation of angular momentum between spins and lattice/phonons which has transpired as the Einstein de-Haas [1] and Barnett effects [2]. This coupling potentially plays a consequential role in the motion of domain walls [3], the polarization of phonons [4], magnetic switching in a mechanical nano-cantilever [3], superconductivity [5], spin relaxation in molecular magnets [6], negative thermal expansion [7] etc. At the same time, light-matter interaction at ultrafast timescales in dimensionally confined systems is an intriguing subject for condensed matter research with enormous application potential [8] and exciting unconventional physics to be uncovered [9]. With the development of ultrashort laser pulses, this manipulation could be achieved and detected at ultrafast timescales which likely results in the all-optical magnetic switching [10]. Magnetic van-der Waals systems intrinsically show strong coupling between different degrees of freedom including spin and lattice projecting [11] themselves as an ideal platform to understand the underlying mechanisms. Hence, it becomes vital to understand the microscopic mechanism for the spin-phonon interaction at its intrinsic timescales.

CrI$_3$ is a prototypical semiconducting van-der Waals ferromagnet with a magnetic ordering temperature of T$_c$=61 K in the bulk [12]. It was one of the first materials where ferromagnetic order was shown to be persistent down to the monolayer limit [13]. The electronic structure of CrI$_3$ is characterized by excitonic and charge transfer transitions [14,15] with strong electron-phonon [16] and magneto-optical coupling [13,14]. The individual layers of $CrI_3$ consist of a hexagonal lattice with $Cr^{3+}$ ions being surrounded by I ligands in an octahedral geometry [14]. The Cr magnetic moment has been determined to be 3 $\mu_B$ with an out-of-plane magnetic orientation [17]. The magnetic interaction in $CrI_3$ is attributed to the superexchange between $Cr^{3+}$ ions mediated via the intermediate I ligands [17] of the Cr-I-Cr bond. This makes the magnetic exchange particularly susceptible to Cr-I atomic displacements as they happen, for instance, in optical phonon modes [18]. $CrI_3$ exhibits a broad range of optical phonon modes in the THz frequency range, with many of being influenced by the onset of magnetic order below $T_c$ [19–21]. Raman scattering has been used to probe phonon modes with 2.4 and 3.9 THz frequencies [18,22]. Satellite peaks due to these phonons have been observed with polarized Raman scattering and were initially interpreted as magnon modes [22] but subsequently shown to be caused by backfolded phonons [18]. Optically induced exciton formation was observed to lead to coherent excitation of the 2.4 and 3.9 THz phonon modes [15,23]. Using optical pump-probe experiments, it was demonstrated that helicity-dependent optical excitation can affect the phonon mode population. However, only the 3.9 THz mode was reported to couple to the magnetic order [23].

In this article, we report the electron and spin dynamics in $CrI_3$ following exciton formation for optical excitation with 1.57 eV photons. The optical generation of excitons in $CrI_3$ is accompanied by a Jahn Teller-like lattice distortion that excites displacive Raman-active phonon modes of $A_{1g}$ and $A_{2g}$ symmetry at 2.4 and 3.9 THz, respectively [15]. We show, using the magneto-optical Kerr effect (MOKE), that both phonon modes couple to the magnetic order, contrary to previous observations [23].

## 2. Results and discussion

We used time-resolved optical and magneto-optical Kerr spectroscopy to investigate the charge and spin dynamics in $CrI_3$. Bulk $CrI_3$ crystals were exfoliated to prepare bulk-like $CrI_3$ flakes similar to ref. [23] which were then transferred under argon atmosphere into a liquid helium (LHe) cryostat. In the cryostat measurements were performed at a pressure of $\sim 10^{-6}$ mbar, circumventing the corrosion of the air-sensitive $CrI_3$ flakes. We used pump laser pulses of 1.57 eV photon energy to excite the sample while time delayed pulses of 2.40 eV probed the dynamics. The pump and probe laser pulses were focused onto the $CrI_3$ flakes of $\sim 70$ μm size to diameters of 40 and 10 μm, respectively. The incidence angle of the p-polarized pump and p-polarized probe beams were $\sim 0°$ and $\sim 22°$ relative to the sample normal, respectively. The overall temporal resolution given by the convolution of pump and probe pulses was 200±5 fs. Measurements were performed at a sample temperature of 4 K under applied out-of-plane magnetic fields of ±0.3 T. The pump-probe time delay traces shown here, are 5 ps long with a step size of 25 fs. Time-resolved changes in reflectivity and MOKE rotation are determined by the average change of the reflected probe intensity and the difference in polarization rotation for two opposite magnetic field directions (±H), respectively. The transient change in reflectivity and magnetism are normalized to the unpumped static values of the corresponding parameters.

Figure 1 shows the measured transient changes in the reflectivity, ΔR/R, and MOKE rotation (ΔM/M) vs. pump-probe time delay (symbols) for different pump fluences. The optical reflectivity signal, ΔR/R, measures changes in the electronic properties of the sample while the MOKE rotation, ΔM/M, is related to the magnetic order. Both ΔR/R and ΔM/M display an initial decrease followed by a slow recovery in the ΔR/R -channel and a further decrease in the ΔM/M-channel. The dynamics after the initial decrease is accompanied by coherent oscillations. In order to extract the size of the time, t, dependence of the coherent oscillations, we model the initial signal drop and further dynamics averaged over the coherent oscillation periods by fitting the data to

$$\frac{\Delta Z}{Z}(t) = A(t, \tau_{fwhm})\, \Delta Z_{Ex} \left[1 + \Delta Z_{ps}\left(1 - e^{-\frac{(t-t_0)}{\tau_Z}}\right)\right] \quad \text{Eq. (1a)}$$

$$\text{with } A(t, \tau_{fwhm}) = g(t, \tau_{fwhm}) \otimes S(t, t_0) \quad \text{and} \quad S(t, t_0) = \begin{cases} 0, & t < t_0 \\ 1, & t \geq t_0 \end{cases} \quad \text{Eq. (1b)}$$

In Eq. (1) $Z = R$ or $M$ describe the reflectivity and magnetism data, respectively, while $\otimes$ denotes the normalized temporal convolution of a step function $S(t, t_0)$ at time-zero, $t_0$, between pump and probe pulses with the experimental temporal resolution function $g(t, \tau_{fwhm})$, modeled as a Gaussian with a full width at half maximum of $\tau_{fwhm}$. The initial signal drop due to exciton formation is given by, $\Delta Z_{Ex}$, in Eq. (1). The subsequent ps dynamics is modeled by an exponential recovery in R and a decay in M, with amplitudes, $\Delta Z_{ps}$ and timescales, $\tau_Z$, for $Z = R, M$. All fit parameters and 1σ error values are summarized in Table 1 for reflectivity, R, and in Table 2 for magnetism, M.

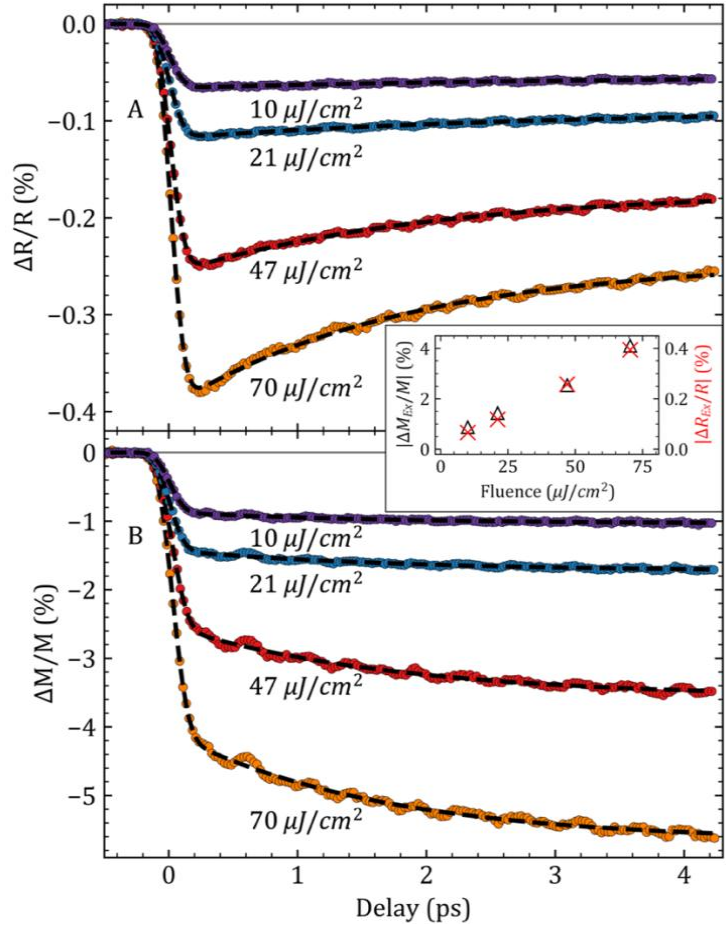

**Figure 1** Ultrafast reflectivity (A) and magnetization dynamics (B) of CrI$_3$ for the pump fluences stated in the figure. Symbols are the experimental data and dashed lines describe the fit according to Eq. (1) as discussed in the text. The inset displays the fluence dependence of the absolute values of the initial signal drop in R (red crosses) and M (black open triangles) channels. The fit parameters are summarized in Tables 1 and 2 for R and M channels, respectively.

The initial signal drops, $\Delta Z_{Ex}$, observed in Fig. 1 show a temporal width that is independent of pump fluence within our temporal resolution. From Tables 1 and 2, we extract an average width of $\tau_{fwhm}$ = 211 ± 10 fs, which is in good agreement with the temporal resolution of our experiment. We use the fitted maxima of A(t, $\tau_{fwhm}$) in Eq. (1) to determine the time-zero, $t_0$, overlap of pump and probe pulses in Fig. 1. The x-axes in all figures are corrected by setting $t_0$ to zero. The size of the initial drop, $\Delta Z_{Ex}$, scales linearly with pump fluence as shown in the inset of Fig. 1. This is the expected behavior for exciton generation since the same amount of photons is required to generate each exciton and increasing the pump fluence will only increase the exciton density. The linear fluence behavior also indicates that we are far below fluences where exciton-exciton interaction can lead to non-linearities [15].

Following exciton formation, the spectra in Fig. 1 display the appearance of coherent oscillations, that will be described below, and further variations of the averaged Z = R, M signals described by the fit parameters $\Delta Z_{ps}$ and $\tau_Z$ in Tables 1, 2. We will discuss the latter first. We note that a continuing decay of M after the initial drop in Fig. 1 was also observed in ref. [23] and was assigned to demagnetization on the ps timescale within a so-called three-temperature model. The ps increase in R in Fig. 1 was reported in ref. [24] and assigned to electronic heating following exciton formation. While the identification of demagnetization and electronic/phononic heating processes are beyond the scope of the present study we can draw

**Table 1.** Fit parameters for reflectivity data as estimated using Eq. (1).

| Fluence (μJ/cm²) | $\Delta R_{Ex}$ (%) | $\Delta R_{ps}$ (%) | $\tau_R$ (ps) | $\tau_{fwhm}$ (ps) |
|---|---|---|---|---|
| 70 | -0.394±0.001 | -0.379±0.002 | 1.8±0.1 | 0.197±0.01 |
| 47 | -0.258±0.001 | -0.332±0.003 | 2.0±0.1 | 0.204±0.008 |
| 21 | -0.118±0.001 | -0.246±0.007 | 2.9±0.2 | 0.208±0.01 |
| 10 | -0.066±0.001 | -0.196±0.016 | 3.3±0.6 | 0.210±0.011 |

**Table 2.** Fit parameters for magnetism data as estimated using Eq. (1).

| Fluence (μJ/cm²) | $\Delta M_{Ex}$ (%) | $\Delta M_{ps}$ (%) | $\tau_M$ (ps) | $\tau_{fwhm}$ (ps) |
|---|---|---|---|---|
| 70 | -4.099±0.02 | 0.383±0.005 | 1.6±0.1 | 0.222±0.012 |
| 47 | -2.539±0.016 | 0.406±0.007 | 1.7±0.1 | 0.225±0.01 |
| 21 | -1.428±0.007 | 0.214±0.005 | 1.8±0.1 | 0.213±0.01 |
| 10 | -0.871±0.005 | 0.192±0.006 | 1.7±0.2 | 0.209±0.011 |

interesting conclusions from the observed fluence dependence.

The values of $\Delta Z_{ps}$ in Eq. 1 represent quantities normalized to the number of generated excitons. $\Delta Z_{ps}$ and $\tau_Z$ should, therefore, be constant with increasing pump fluence which is clearly not the case in Tables 1, 2. We assign this deviation to polaronic regions developing around each exciton where electronic structure and magnetic order is affected by the non-equilibrium exciton formation. With increasing pump fluence, both $\Delta R_{ps}$ and $\Delta M_{ps}$ increase to larger absolute values. This indicates that the polaronic regions around excitons start to overlap and influence each other. We note that this behavior is not observed for the exciton formation itself (see inset of Fig. 1) likely because the formed excitons are thought to be of atomic-like

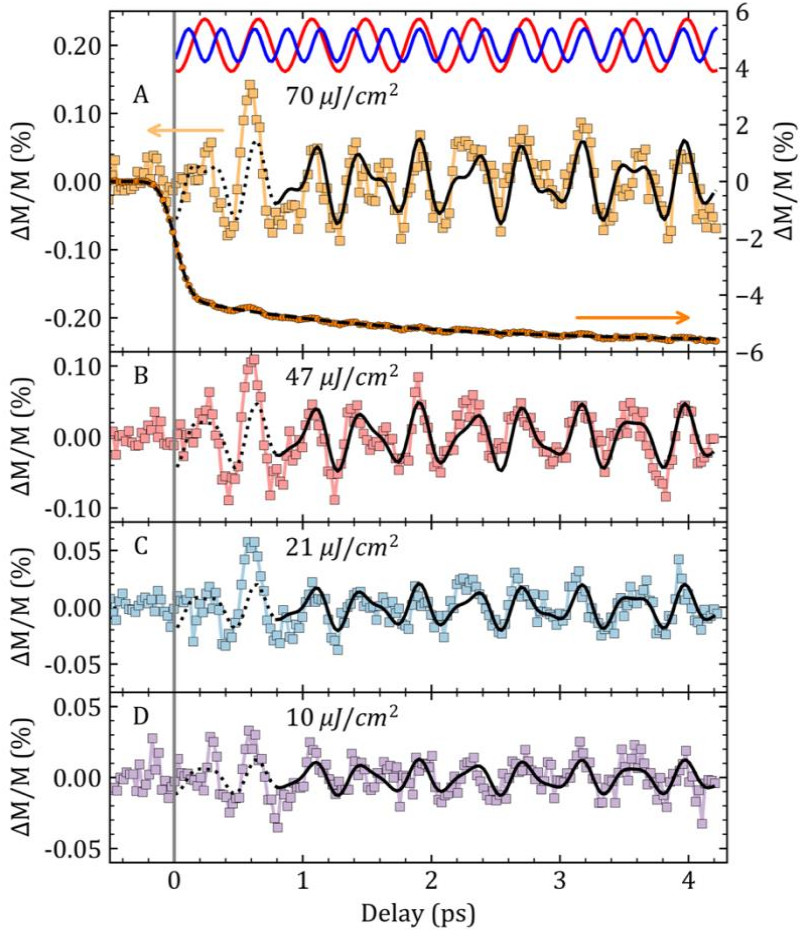

**Figure 2** Coherent oscillations in the magnetic signal for different fluences. The oscillatory component of the magnetic signal is retrieved by removing the smooth background according to Eq. (1) from the measured ΔM/M signals, representatively shown in (A). The red and blue oscillations in (A) are the individual contributions of the 2.4 and 3.9 THz modes to the coherent oscillations in the magnetic signal. Experimental ΔM/M values due to the coherent oscillations are shown as colored lines and symbols. Fits according to Eq. (2) are shown as black solid lines, while the dotted black line is the extrapolation to t=0 for the visualization of the phase. The fit parameters are summarized in Table 3.

extend [13] and possibly too small to influence each other.

After the initial signal drop in Fig. 1, coherent oscillations are observed in ΔM/M and to a lesser degree in ΔR/R. Figure 2 displays the results for coherent oscillations visible after subtracting the background ΔM/M signal according to Eq. (1) while Fig. 3 shows the coherent oscillations for ΔR/R. We fitted the coherent oscillations in Z = R and M as

$$f_Z(t) = A_{Z,2.4THz} \sin(2\pi \nu_{Z,2.4THz}\, t + \phi_{Z,2.4THz}) + A_{Z,3.9THz} \sin(2\pi \nu_{Z,3.9THz}\, t + \phi_{Z,3.9THz})$$
Eq. 2

where $A$, $\nu$ and $\phi$ describe amplitudes, frequencies and phases for the 2.4 THz and 3.9 THz modes. Here we made the simplifying assumption that only two modes contribute to the oscillations and that their frequencies are identical to 2.4 THz and 3.9 THz as reported previously [15,23,24]. The resulting fit parameters and 1σ error values are summarized in

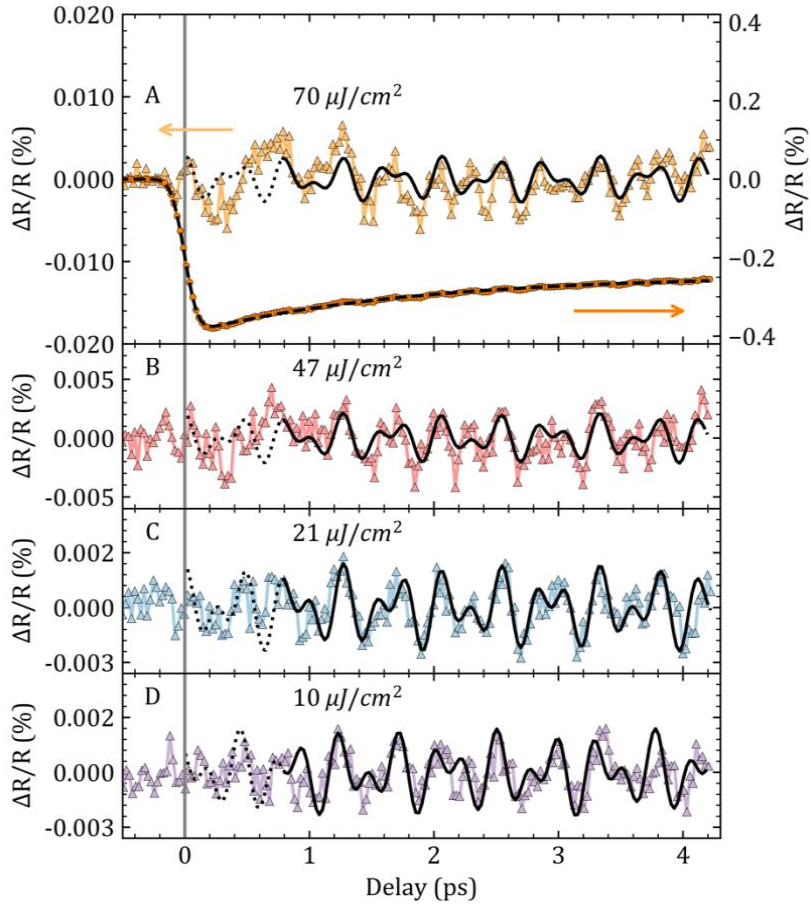

**Figure 3** Coherent oscillations in the reflectivity signal for different fluences same as in fig. 2.

Tables 3 and 4 for M and R, respectively.

The coherent oscillations visible in the magnetic signal can be reasonably well described for all pump fluences by the contributions of 2.4 and 3.9 THz modes for delay times longer than about 1 ps (see Fig. 2). For times earlier than 1 ps, deviations for the sinusoidal fits (dashed lines) are seen consistently for all fluences. For instance, a significantly larger positive oscillation is seen around 0.5 ps. This could indicate that it takes some time for the oscillations to form a steady state behavior. The obtained steady-state amplitudes scale linearly with the pump fluence as can be seen in Fig. 4A. This is in agreement with the notion that it is the excitons that drive the observed oscillations and their amplitudes are dependent on the formation mechanism which depends on the nature of the exciton but not their density. Once started the magnetization oscillations would continue and their all-over amplitude would mainly scale with the exciton density and be affected to a lesser degree by further demagnetization.

Figure 3 shows the background-corrected reflectivity oscillations of the data in Fig. 1. It is apparent that the oscillation signal is less pronounced in reflectivity compared to that in the magnetization. While for some delay time ranges oscillations are clearly seen, they appear less pronounced for others. We fitted the data in Fig. 3 similarly to the procedure described above for the magnetism oscillations. It is however noteworthy that we can obtain a better fit by relaxing the requirement to keep the two oscillation frequencies fixed. This is another indication that the oscillations in reflectivity (Fig.3) are much closer to the noise level that the corresponding ones in magnetism (Fig. 2).

Figure 4 displays the fluence dependence of both phonon amplitudes in magnetism and reflectivity. They appear to increase linearly with fluence for the magnetism channel. This indicates that their strength scales with the number of generated excitons. On the other hand, the extracted amplitude values for the oscillations in reflectivity are too close to the noise level to reach to a proper conclusion for their fluence dependence. We note that in this data analysis we keep the frequencies fixed to the literature values of 2.4 and 3.9 THz [15,23,24], while only amplitudes and phases for both modes are free parameters and listed in Tables 3 and 4 for magnetism and reflectivity channels.

**Table 3.** Magnetization amplitudes, $A_M$, and phases, $\phi_M$, of two modes with frequencies of 2.4 and 3.9 THz for different fluences.

| Fluence (µJ/cm$^2$) | $A_{M,2.4THz}$ (%) | $\phi_{M,2.4THz}$ (rad) | $A_{M,3.9THz}$ (%) | $\phi_{M,3.9THz}$ (rad) |
|---|---|---|---|---|
| 70 | 0.039±0.003 | -2.0±0.1 | 0.024±0.003 | -1.2±0.1 |
| 47 | 0.035±0.003 | -1.7±0.1 | 0.015±0.003 | -0.9±0.2 |
| 21 | 0.015±0.001 | -1.8±0.1 | 0.008±0.001 | -0.7±0.2 |
| 10 | 0.009±0.001 | -1.9±0.1 | 0.004±0.001 | -1.1±0.3 |

**Table 4.** Reflectivity amplitudes, $A_R$, and phases, $\phi_R$, of two modes with frequencies of 2.4 and 3.9 THz for different fluences.

| Fluence (µJ/cm$^2$) | $A_{R,2.4THz}$ (%) | $\phi_{R,2.4THz}$ (rad) | $A_{R,3.9THz}$ (%) | $\phi_{R,3.9THz}$ (rad) |
|---|---|---|---|---|
| 70 | 0.0016±0.0003 | 1.7±0.2 | 0.0012±0.0003 | 1.3±0.2 |
| 47 | 0.0011±0.0003 | 1.4±0.1 | 0.0011±0.0003 | 1.7±0.2 |
| 21 | 0.0012±0.0001 | 1.2±0.1 | 0.001±0.0001 | 1.7±0.1 |
| 10 | 0.0010±0.0001 | 1.3±0.1 | 0.001±0.0001 | -3.1±0.1 |

Similar oscillations have been observed in ref. [15] for reflectivity above the magnetic ordering temperature and in ref. [24] below $T_c$. The Fourier transforms of the coherent phonon oscillations in refs. [15, 23, 24] show two peaks at 2.4 and 3.9 THz. However, the oscillations in the magnetism channel of ref. [23] were only observed for a frequency of 3.9 THz. Contrary to ref. [23], our results clearly show that both phonon modes couple to magnetism. One explanation got this discrepancy could be the detection method. In this work, we use the conventional scheme of time-resolved MOKE where the sample is magnetically saturated by an external magnetic field along opposite directions. The difference is MOKE signal is then directly related to the laser induced spin dynamics [25]. In ref. [23] the magnetic sensitivity was instead induced by pumping with circularly polarized radiation while keeping the sample free of any magnetic field. It will be interesting to see systematic future studies addressing this behavior.

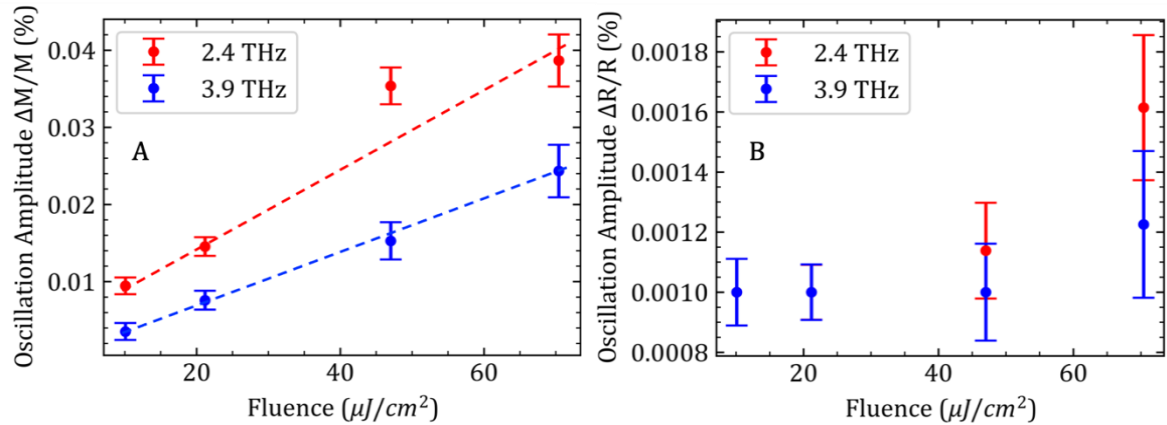

**Figure 4** Fluence dependence of the amplitude of coherent vibrational modes. A) and B) The amplitude of 2.4 (red markers) and 3.9 (blue markers) THz in magnetic and reflectivity channel, respectively. The amplitude of both modes in M increases linearly with fluence as indicated by the red and blue dashed lines.

### 3. Conclusion
We showed that the ultrafast generation of excitons induces spin and phonon dynamics in the semiconducting 2D ferromagnet $CrI_3$. Two coherent phonon modes have been identified as the bending and stretching of the Cr-I bonds at frequencies of 2.4 and 3.9 THz, respectively, in agreement with previous studies [15,23,24]. Contrary to previous investigations [23] we find that both phonon modes coupled to magnetism, i.e., we identify the same coherent oscillation in a time-resolved MOKE signal. In addition, we find that the exciton formation leads to the formation of extended polaronic regions that also affect the magnetic order. Contrary to the coherent spin and phonon motion, these regions start to influence each other with increasing exciton density.

### Acknowledgement
V.S., M.P., J.R. and H.A.D. acknowledge support by the Swedish Research Council (VR) and the Knut and Alice Wallenberg Foundation (KAW). D.M.B. and M.V.K. gratefully acknowledge funding from the European Research Council (ERC) Project SPINNER (Grant No. 101002772) and Knut and Alice Wallenberg Foundation (Grants No. 2022.0079 and 2023.0336).